\documentclass[aps,plb,preprint,superscriptaddress]{revtex4}

\usepackage{youngtab}
\usepackage{epsfig}

\begin{document}
\title{\boldmath The $qqqq\bar{q}$ components and 
hidden flavor contributions to the baryon magnetic moments}

\author{C. S. An}
\email[]{ancs@mail.ihep.ac.cn}
\affiliation{Institute of High Energy Physics, CAS, P. O. Box 918,
Beijing 100049, China}

\author{Q. B. Li}
\email[]{ligb@pcu.helsinki.fi}
\affiliation{Helsinki Institute of Physics, P. O. Box 64,00014
University of Helsinki, Finland}

\author{D. O. Riska}
\email[]{riska@pcu.helsinki.fi}
\affiliation{Helsinki Institute of Physics, P. O. Box 64,00014 
University of Helsinki, Finland}

\author{B. S. Zou}
\email[]{zoubs@mail.ihep.ac.cn}
\affiliation{CCAST (World Lab.), P. O. Box 8730, Beijing 100080,
China}
\affiliation{Institute of High Energy Physics, CAS, P. O. Box 918,
Beijing 100049, China}

\date{\today}

\begin{abstract}
The contributions from the $qqqq\bar q$ components
to the magnetic moments of the octet as well as the $\Delta^{++}$
and $\Omega^-$ decuplet baryons are calculated for the
configurations that are expected to have the lowest
energy if the hyperfine interaction depends on both
spin and flavor. The contributions from
the $u\bar u$, $d\bar d$ and the $s\bar s$
components are given separately. 
It is shown that addition of $qqqq\bar{q}$ admixtures to the
ground state baryons can improve the overall description of
the magnetic moments of the baryon octet and decuplet in the quark
model without SU(3) flavor symmetry  breaking, beyond that
of the different constituent masses of the strange
and light-flavor quarks. 
The explicit flavor (and spin) wave functions for all the possible 
configurations
of the $qqqq\bar{q}$ components
with light and strange $q\bar q$ pairs are given for the baryon 
octet and decuplet.
Admixtures of $\sim$ 10\% of the $qqqq\bar q$ configuration
where the flavor-spin symmetry is $[4]_{FS}[22]_F[22]_S$, which
is likely to have the lowest energy, in particular reduces the 
deviation from the empirical values of the magnetic moments 
$\Sigma^-$ and the $\Xi^0$ compared with 
the static $qqq$ quark model. 
\end{abstract}

\maketitle

\section{introduction}
\label{sec:1}

Recent measurements of the
$\bar d/\bar u$ asymmetry in the nucleon sea
indicate a considerable isospin symmetry breaking in the light quark
sea of nucleon. This indicates that the nucleon contains notable
$qqqq\bar q$ components, if not more exotic
components, besides the conventional $qqq$ component 
\cite{towell,NA51,NMC,HERMES}. The experiments on parity
violation in electron-proton scattering moreover indicate that
$s\bar s$ quark pairs lead to nonzero contributions to the magnetic 
moment of the proton\cite{happex1,happex2,sample1,sample2,a41,a42,g0}. 

Here the contributions to the baryon magnetic
moments from wave function components with at most one $u\bar u$, $d\bar d$
or $s\bar s$ sea quark pair are calculated in the non-relativistic
quark model. The conventional $qqq$ constituent quark model by
itself provides a qualitative description
of the magnetic moments of the baryon octet. Quantitatively the
description is, however,  not better than $\sim$ 15 \%, the
largest differences from the experimental values being the
magnetic moments of the $\Sigma^-$ and the $\Xi^0$. While
covariant reformulation of the $qqq$ model does not change
this situation \cite{bruno}, the overall
description may be improved somewhat by adding effects of
meson exchange and orbital excitations to the $qqq$ model
wave functions \cite{Franklin1, Franklin2, dannb}. This of 
course also suggests that
explicit $q\bar q$ terms should be included in the quark
model for an improved description. 
Such explicit $qqqq\bar q$ components with light $q\bar q$ pairs have
been shown to
improve significantly the agreement between the calculated and
the measured the decays of low lying baryon resonances
\cite{delpion,delgamma,roper}.

The experimental results on the strangeness magnetic moments
can be described, at least qualitatively,
by $uuds\bar s$ configurations in the proton, where the
$\bar s$ antiquark is in the $S-$state  \cite{zou1,zou2,zou3}.
Since configurations with the antiquark in the $S-$state cannot
be represented by long range pion or kaon loop fluctuations, this
motivates to a systematic extension of the $qqq$ 
quark model to include the $qqqq\bar q$ configurations explicitly.

Here the explicit flavor and spin wave functions for all the possible
configurations of the $qqqq\bar{q}$ system in the baryon octet and
$\Delta^{++}$ and $\Omega^{-}$ 
decuplet baryons, with
totally symmetric spin-flavor symmetry are derived, when both light 
and strange $q\bar q$ pairs are taken into account. Finally the
contributions to the magnetic moments of the octet and
decuplet baryons are
evaluated for the mixed symmetry configurations $[4]_{FS}[22]_F[22]_S$
and  $[4]_{FS}[31]_F[31]_S$,
which are likely to have the lowest energy in the case of the
octet and decuplet baryons, respectively. This notation is a
shorthand for the Young tableaux decomposition:
\begin{eqnarray}
&&[4]_{FS}[22]_F[22]_S\,:\,\Yvcentermath1
{\yng(4)}_{FS}\,{\yng(2,2)}_F\,{\yng(2,2)}_S \,, 
\nonumber\\
&&[4]_{FS}[31]_F[31]_S\,:\,\Yvcentermath1
{\yng(4)}_{FS}\,{\yng(3,1)}_F\,{\yng(3,1)}_S \, .
\end{eqnarray}

An interesting result is that
this extension of the quark model allows to a notable reduction
of the overall disagreement with the empirical magnetic moments
when the $qqqq\bar q$ admixture is of the order of $\sim$ 10\% or
more and if the $qqqq\bar q$ component is more compact than
the $qqq$ component. The improvement is particularly notable in the 
case of the $\Sigma^{-}$ and $\Xi^{0}$ hyperons.

Explicit expressions for the magnetic moments 
from the $u\bar u$, $d\bar d$ as well as well as the 
$s\bar s$ seaquark pairs  
are given.
From these the ``strangeness magnetic'' moments - i.e. the
magnetic moment contributions from $s\bar s$ pairs -  of all the baryons
can be read off. In addition the light flavor seaquark
contributions as e.g. the contributions 
of $u\bar u$ and $d\bar d$ pairs to the magnetic moment of the $\Omega^-$
are given explicitly. 

The present paper is organized in the following way.
In Section \ref{sec:2} the explicit flavor
wave functions of the $qqqq\bar{q}$ components in the baryon octet and
decuplet from the $SU(3)$ symmetry are given. Section \ref{sec:3} contains the 
expressions 
and the corresponding numerical results for the baryon octet.
The corresponding expressions for the  
2 decuplet magnetic moments are given in section \ref{sec:4}.
The strangeness magnetic moments are considered in \ref{sec:5} and
the light flavor seaquark magnetic moments in \ref{sec:6}.
Finally section \ref{sec:7} contains a concluding discussion.

\section{the flavor wave functions of the $qqqq\bar{q}$ components}
\label{sec:2}

The color symmetry of the $qqqq$
subsystem in a $qqqq\bar{q}$
component in a baryon is limited to $[211]_{C}$ by the requirement
that it form a
color singlet in combination with
the anti-quark. The Pauli principle then requires that
the corresponding orbital-flavor-spin states have the  
mixed symmetry $[31]_{XFS}$ in order to combine with the color state 
$[211]_{C}$ 
to the required completely
antisymmetric 4q state $[1111]$. 
Since the intrinsic parity is
positive for a quark and negative for an anti-quark, the
$qqqq\bar{q}$ components, which have positive parity, require
that either the $\bar{q}$ is in the P-state and that the $qqqq$
subsystem is in the spatially symmetric
ground state ($[4]_{X}$) or
that one of the quarks is in the P-state so that the $qqqq$
subsystem has mixed spatial symmetry $[31]_{X}$ and that
the $\bar q$ is in its ground state. 
The extant experimental data on the strangeness magnetic moment of the
proton suggests that the $qqqq$ subsystem has mixed spatial  
symmetry $[31]_X$ if one of the quarks is strange \cite{zou1,zou2,zou3}. 

The possible spin symmetry states of four quarks are: $[4]_{S}$,
$[31]_{S}$ and $[22]_{S}$. There are four possible flavor symmetry
configurations for the $qqqq$ subsystem, which can combined with
the spatial and spin symmetry to form the orbital-flavor-spin
symmetry $[31]_{XFS}$: $[4]_{F}$, $[31]_{F}$, $[22]_{F}$ and
$[211]_{F}$ in the Weyl tableaux of the group $SU(3)$ \cite{ma,chen}. 
Combination of these flavor states with
the anti-quark with flavor $[1]^{*}_{F}$ leads to the following
$qqqq\bar{q}$ multiplet representations of $SU(3)$:
 \begin{equation}
 [4]_{F}\otimes[1]^{*}=\bf{10}\oplus\bf{35}\, ,
\end{equation}
\begin{equation}
 [31]_{F}\otimes[1]^{*}=\bf{8}\oplus\bf{10}\oplus\bf{27}\, ,
\end{equation}
\begin{equation}
[22]_{F}\otimes[1]^{*}=\bf{8}\oplus\bf{\bar{10}}\, ,
\end{equation}
\begin{equation}
 [211]_{F}\otimes[1]^{*}=\bf{1}\oplus\bf{8}\, .
\end{equation}
Here the numbers in boldface on the right-hand-side of the
equations indicate the dimensions of the pentaquark
representations. As an example the possible
$\theta^{+}$-pentaquark may belong to the baryon anti-decuplet
$\bar{\textbf{10}}$ representation \cite{diak}. It's obvious that the
$qqqq\bar{q}$ components of the baryon decuplet belong to the
$\textbf{10}$ representations, and that of the baryon octet
belongs to the $\textbf{8}$ representations. The combination of
the flavor states with the spin states gives rise to several
flavor-spin states, which can be split by the spin-dependent
hyperfine interaction between the quarks. 

If the hyperfine interaction between the quarks depends on spin- and
flavor \cite{glozman} the $qqqq$ systems with the mixed spatial
symmetry $[31]_X$ are expected to be the configurations with the lowest
energy, and therefore most likely to form appreciable 
components of the proton. Consequently the flavor-spin
state of the $qqqq$ system is most likely totally symmetric: $[4]_{FS}$. 
Moreover in the case of the baryon octet the flavor-spin state 
$[4]_{FS}[22]_{F}[22]_{S}$, with one quark in
its first orbitally excited state, is likely to have the lowest 
energy \cite{hr}.
For the baryon decuplet the flavor symmetry $[22]_{F}$ is, however, not
possible for the $qqqq$ subsystem,
and the $[4]_{FS}[31]_{F}[31]_{S}$ symmetry configuration
is expected to have the lowest energy \cite{hr}.

A baryon wave function that includes
$qqqq\bar q$ components in addition to
the conventional $qqq$ components may be written
in the following general form:
\begin{equation}
|B\rangle=\sqrt{P_{3q}}|qqq\rangle+\sqrt{P_{5q}}\sum_{i}A_{i}|qqqq_{i}
\bar{q_{i}}\rangle \, .
\label{bwave}
\end{equation}
Here $P_{3q}$ and $P_{5q}$ are the probabilities of the $qqq$ and $qqqq\bar{q}$
components respectively; the sum over i runs over all the possible
$qqqq_{i}\bar{q_{i}}$ components, and $A_{i}$ denotes the amplitude of the
corresponding 5q component. The wave functions of the $qqq$ components are the
conventional $SU(6)_{SF}$ ones.
Here  the flavor (and spin) wave
functions of every $qqqq_{i}\bar{q_{i}}$ component is constructed along
with a calculation of the
corresponding amplitudes $A_{i}$. Note that the baryons still
have the $SU(3)$  flavor symmetry when the $qqqq\bar{q}$ 
components have been taken
into account.

The weight diagram method \cite{ma} of the SU(3) group will be employed
here for the explicit construction of the wave functions of baryon multiplet
in $qqqq_{i}\bar{q_{i}}$ configurations. These wave functions
and amplitudes for the baryon octet and decuplet, which have complete 
flavor-spin
symmetry $[4]_{FS}$
are listed in tables
\ref{table1}, \ref{table2} and \ref{table3}.

\begin{table}[h]\caption{\footnotesize The $qqqq\bar{q}$ components, in which 
the
$qqqq$ subsystem has the flavor symmetry $[4]_{F}$.}
\vspace{0.5cm}
\footnotesize
\begin{tabular}{l|l}
\hline
($\frac{3}{2}^{+},Y,I,I_{3}$)& $qqqq\bar{q}$ component\\\hline
(1,3/2,+3/2)& $\sqrt{\frac{2}{3}}|uuuu\bar{u}\rangle+\sqrt{\frac{1}{6}}
|uuud\bar{d}\rangle+\sqrt{\frac{1}{6}}|uuus\bar{s}\rangle$\\
(1,3/2,+1/2)& $\sqrt{\frac{1}{2}}|uudu\bar{u}\rangle+\sqrt{\frac{1}{3}}
|uudd\bar{d}\rangle+\sqrt{\frac{1}{6}}|uuds\bar{s}\rangle$\\
(1,3/2,-1/2)& $\sqrt{\frac{1}{3}}|uddu\bar{u}\rangle+\sqrt{\frac{1}{2}}
|uddd\bar{d}\rangle+\sqrt{\frac{1}{6}}|udds\bar{s}\rangle$\\
(1,3/2,-3/2)& $\sqrt{\frac{1}{6}}|dddu\bar{u}\rangle+\sqrt{\frac{2}{3}}
|dddd\bar{d}\rangle+\sqrt{\frac{1}{6}}|ddds\bar{s}\rangle$\\
(0,1,+1)&     $\sqrt{\frac{1}{2}}|uusu\bar{u}\rangle+\sqrt{\frac{1}{6}}
|uusd\bar{d}\rangle+\sqrt{\frac{1}{3}}|uuss\bar{s}\rangle$\\
(0,1,0)&     $\sqrt{\frac{1}{3}}|udsu\bar{u}\rangle+\sqrt{\frac{1}{3}}
|udsd\bar{d}\rangle+\sqrt{\frac{1}{3}}|udss\bar{s}\rangle$\\
(0,1,-1)&     $\sqrt{\frac{1}{6}}|ddsu\bar{u}\rangle+\sqrt{\frac{1}{2}}
|ddsd\bar{d}\rangle+\sqrt{\frac{1}{3}}|ddss\bar{s}\rangle$\\
(-1,1/2,+1/2)&     $\sqrt{\frac{1}{3}}|ussu\bar{u}\rangle+\sqrt{\frac{1}{6}}
|ussd\bar{d}\rangle+\sqrt{\frac{1}{2}}|usss\bar{s}\rangle$\\
(-1,1/2,-1/2)&     $\sqrt{\frac{1}{6}}|dssu\bar{u}\rangle+\sqrt{\frac{1}{3}}
|dssd\bar{d}\rangle+\sqrt{\frac{1}{2}}|dsss\bar{s}\rangle$\\
(-2,0,0)&     $\sqrt{\frac{1}{6}}|sssu\bar{u}\rangle+\sqrt{\frac{1}{6}}
|sssd\bar{d}\rangle+\sqrt{\frac{2}{3}}|ssss\bar{s}\rangle$\\
\hline
\end{tabular}
\label{table1}
\end{table}

\begin{table}[h]\caption{\footnotesize The $qqqq\bar{q}$ components, in which 
the
$qqqq$ subsystem has the flavor symmetry $[31]_{F}$.}
\vspace{0.5cm}
\footnotesize
\begin{tabular}{l|l|l|l}
\hline
($\frac{3}{2}^{+},Y,I,I_{3}$)& $qqqq\bar{q}$ component&($\frac{1}{2}^{+},
Y,I,I_{3}$)& $qqqq\bar{q}$ component\\\hline
(1,3/2,+3/2)& $-\sqrt{\frac{1}{2}}(|uuud\bar{d}\rangle+|uuus\bar{s}\rangle)$
&&\\
(1,3/2,+1/2)&$\sqrt{\frac{1}{6}}|uudu\bar{u}\rangle-\sqrt{\frac{1}{3}}
|uudd\bar{d}\rangle$&
(1,1/2,+1/2)&$-(\sqrt{\frac{8}{15}}|uudu\bar{u}\rangle+\sqrt{\frac{4}{15}}
|uudd\bar{d}\rangle$\\
&$-\sqrt{\frac{1}{2}}|uuds\bar{s}\rangle$ &&
$+\sqrt{\frac{3}{15}}|uuds\bar{s})\rangle$\\
(1,3/2,-1/2)& $\sqrt{\frac{1}{3}}|uddu\bar{u}\rangle-\sqrt{\frac{1}{6}}
|uddd\bar{d}\rangle$&
(1,1/2,-1/2)&$-(\sqrt{\frac{4}{15}}|uddu\bar{u}\rangle+\sqrt{\frac{8}{15}}
|uddd\bar{d}\rangle$\\
&$-\sqrt{\frac{1}{2}}|udds\bar{s}\rangle$
&&$+\sqrt{\frac{3}{15}}|udds\bar{s}\rangle)$\\
(1,3/2,-3/2)& $\sqrt{\frac{1}{2}}|dddu\bar{u}\rangle-\sqrt{\frac{1}{2}}
|ddds\bar{s}\rangle$&&\\
(0,1,+1)&     $\sqrt{\frac{1}{6}}|uusu\bar{u}\rangle-\sqrt{\frac{1}{2}}
|uusd\bar{d}\rangle$&
(0,1,+1)&$-(\sqrt{\frac{8}{15}}|uusu\bar{u}\rangle+\sqrt{\frac{3}{15}}
|uusd\bar{d}\rangle$\\
&$-\sqrt{\frac{1}{3}}|uuss\bar{s}\rangle
$&&$+\sqrt{\frac{4}{15}}|uuss\bar{s}\rangle)$\\
(0,1,0)&     $\sqrt{\frac{1}{3}}|udsu\bar{u}\rangle
-\sqrt{\frac{1}{3}}
|udsd\bar{d}\rangle
$&
(0,1,0)&$\sqrt{\frac{11}{30}}|udsu\bar{u}\rangle-\sqrt{\frac{11}{30}}
|udsd\bar{d}\rangle$\\
&$
-\sqrt{\frac{1}{3}}|udss\bar{s}\rangle
$
&&$-\sqrt{\frac{4}{15}}|udss\bar{s}\rangle$\\
(0,1,-1)&     $\sqrt{\frac{1}{2}}|ddsu\bar{u}\rangle+\sqrt{\frac{1}{6}}
|ddsd\bar{d}\rangle
$&
(0,1,-1)&$\sqrt{\frac{3}{15}}|ddsu\bar{u}\rangle-\sqrt{\frac{8}{15}}
|ddsd\bar{d}\rangle$\\
&$
-\sqrt{\frac{1}{3}}|ddss\bar{s}\rangle
$&&$-\sqrt{\frac{4}{15}}|ddss\bar{s}\rangle$\\
(-1,1/2,+1/2)&     $\sqrt{\frac{1}{3}}|ussu\bar{u}\rangle-\sqrt{\frac{1}{2}}
|ussd\bar{d}\rangle
$&
(-1,1/2,+1/2)&$-(\sqrt{\frac{4}{15}}|ussu\bar{u}\rangle+\sqrt{\frac{3}{15}}
|ussd\bar{d}\rangle$\\
&$
-\sqrt{\frac{1}{6}}|usss\bar{s}\rangle
$&&$+\sqrt{\frac{8}{15}}|usss\bar{s}\rangle)$\\
(-1,1/2,-1/2)&     $\sqrt{\frac{1}{2}}|dssu\bar{u}\rangle+\sqrt{\frac{1}{3}}
|dssd\bar{d}\rangle
$&
(-1,1/2,-1/2)&$ \sqrt{\frac{3}{15}}|dssu\bar{u}\rangle-\sqrt{\frac{4}{15}}
|dssd\bar{d}\rangle$\\
&$
+\sqrt{\frac{1}{6}}|dsss\bar{s}\rangle
$&&$-\sqrt{\frac{8}{15}}|dsss\bar{s}\rangle$\\
(-2,0,0)&     $\sqrt{\frac{1}{2}}(|sssu\bar{u}\rangle+|sssd\bar{d}\rangle)$&
(0,0,0)&$-(\sqrt{\frac{3}{10}}|udsu\bar{u}\rangle+\sqrt{\frac{3}{10}}
|udsd\bar{d}\rangle$\\
&&&$+\sqrt{\frac{4}{10}}|udss\bar{s}\rangle)$\\
\hline
\end{tabular}
\label{table2}
\end{table}

\begin{table}[h]\caption{\footnotesize The $qqqq\bar{q}$ components,
 in which the
$qqqq$ subsystem has the flavor symmetry $[22]_{F}$ and $[211]_{F}$.}
\vspace{0.5cm}
\footnotesize
\begin{tabular}{l|l|l}
\hline
($\frac{1}{2}^{+},Y,I,I_{3}$)& $qqqq\bar{q}$ component ($[22]_{F}$)& 
$qqqq\bar{q}$
component ($[211]_{F}$)\\\hline
(1,1/2,+1/2)&$\sqrt{\frac{2}{3}}|uudd\bar{d}\rangle+\sqrt{\frac{1}{3}}
|uuds\bar{s}\rangle$ &$|uuds\bar{s}\rangle$\\
(1,1/2,-1/2)&$\sqrt{\frac{2}{3}}|uddu\bar{u}\rangle-\sqrt{\frac{1}{3}}
|udds\bar{s}\rangle$ &$|udds\bar{s}\rangle$\\
(0,1,+1)&$\sqrt{\frac{1}{3}}|uusd\bar{d}\rangle+\sqrt{\frac{2}{3}}
|uuss\bar{s}\rangle$ &$-|uusd\bar{d}\rangle$\\
(0,1,0)&$-(\sqrt{\frac{1}{6}}|udsu\bar{u}\rangle-\sqrt{\frac{1}{6}}
|udsd\bar{d}\rangle)+\sqrt{\frac{2}{3}}|udss\bar{s}\rangle$&
$\sqrt{\frac{1}{2}}(|udsu\bar{u}\rangle-|udsd\bar{d}\rangle)$\\
(0,1,-1)&$-(\sqrt{\frac{1}{3}}|ddsu\bar{u}\rangle-\sqrt{\frac{2}{3}}
|ddsd\bar{d}\rangle)$ &$|ddsu\bar{u}\rangle$\\
(-1,1/2,+1/2)&$-(\sqrt{\frac{2}{3}}|ussu\bar{u}\rangle-\sqrt{\frac{1}{3}}
|ussd\bar{d}\rangle)$ &$-|ussd\bar{d}\rangle$\\
(-1,1/2,-1/2)&$-(\sqrt{\frac{1}{3}}|dssu\bar{u}\rangle+\sqrt{\frac{2}{3}}
|dssd\bar{d}\rangle)$& $|ddsu\bar{u}\rangle$\\
(0,0,0)&$-\sqrt{\frac{1}{2}}(|udsu\bar{u}\rangle+|udsd\bar{d}\rangle)$&
$\sqrt{\frac{1}{6}}|udsu\bar{u}\rangle-\sqrt{\frac{1}{6}}|udsd\bar{d}
\rangle+\sqrt{\frac{2}{3}}|udss\bar{s}\rangle)$\\
\hline
\end{tabular}
\label{table3}
\end{table}

\section{the baryon octet magnetic moments}
\label{sec:3}

\subsection{Wave functions}

The baryon wave function is formed as combinations of the color, space,
flavor and spin wave functions with appropriate Clebsch-Gordan
coefficients. 
Here the states, in which the
antiquark is in its ground state, so that the flavor-spin 
state of the $qqqq$ system is completely symmetric ($[4]_{FS}$)
are considered. 
The flavor-spin configuration of the $qqqq$ system, which is
expected to have the lowest energy for an octet baryon,
is for the reasons mentioned above, the mixed symmetry
configuration $[4]_{FS}[22]_F[22]_S$. The flavor-spin
decomposition of this wave function is:
\begin{equation}
|[4]_{FS}[22]_F[22]_S \rangle = {1\over \sqrt{2}}\{[22]_{F_1}[22]_{S_1}
+[22]_{F_2}[22]_{S_2}\}\, .
\label{lowesta}
\end{equation}
This expression may be rewritten more pictorially in terms of
Young tableaux as:
\begin{equation}
\Yvcentermath1
{\yng(4)}_{FS}\, {\yng(2,2)}_F\,{\yng(2,2)}_S\,=\,{1\over\sqrt{2}}\bigg(\,
{\young(12,34)}_F\,{\young(12,34)}_S\,+\, 
{\young(13,24)}_F\,{\young(13,24)}_S\,\bigg) \,.
\end{equation}
The explicit forms of the $qqqq\bar{q}$ flavor wave functions, in which the 
$qqqq$ subsystem has flavor symmetry $[22]_F$ are listed
in Table III. 
The corresponding spin wave functions 
are readily
derived from these flavor wave functions by the substitutions:
$u\rightarrow \uparrow$, $ d\rightarrow\downarrow$ and
$ s\rightarrow\downarrow$.

The corresponding flavor-spin configuration of the
decuplet baryons is $[4]_{FS}[31]_F[31]_S$. 
These wave functions take the general form:
\begin{equation}
|[4]_{FS}[31]_F[31]_S \rangle = {1\over \sqrt{3}}\{[31]_{F_1}[31]_{S_1}
+[31]_{F_2}[31]_{S_2}+[31]_{F_3}[31]_{S_3}
\}\, .
\label{lowestb}
\end{equation}
The
explicit forms of the $qqqq\bar{q}$ flavor wave functions, in which the 
$qqqq$ subsystem have flavor symmetry $[31]_F$ are
listed in Table II. The corresponding spin wave functions
are obtained from these flavor wave functions by the same
substitutions as above.

The remaining 19 different symmetric combinations of mixed
symmetry flavor and spin wave functions for the $qqqq$
system are listed in Table 3 of ref. \cite{hr}. The
appropriate $S_4$ Clebsch-Gordan coefficients for the
completely symmetric combinations of these flavor
and spin wave functions are listed in Table 4.13 of ref.
\cite{chen}.

In the non-relativistic quark model, the magnetic moment 
of a baryon is defined
as the expectation value of the magnetic moment operator: 
\begin{equation}
\hat{\mu}=\sum_{i}\frac{Q_{i}}{2m_{i}}(\hat{l}_{iz}+\hat{\sigma}_{iz}).
\label{op}
\end{equation}
Here the sum over i runs over the quark content of the baryon, and $Q_{i}$
denotes the corresponding electric charge of the quark and $m_{i}$ are the 
constituent quark masses.  

With the combination of $qqq$ and $qqqq\bar q$ state wave functions
(\ref{bwave}) the magnetic moment will have
contributions from the diagonal matrix elements of the operator
(\ref{op}) between the $qqq$ component and the $qqqq\bar q$
components, respectively, and from the off-diagonal matrix elements
between the $qqq$ and the $qqqq\bar q$ components. These 
transition matrix elements between the $qqq$ and
$qqqq\bar{q}$ components typically give rise to larger
contributions to the
baryon magnetic moments than the diagonal contributions
from the $qqqq\bar q$ components. The contributions to the
magnetic moment operator from the non-diagonal terms,
which involve $q\bar q$ pair annihilation and creation, are
obtained as matrix elements of the operator:
\begin{equation}
\hat{\mu}=\sum_{i}\frac{Q_{i}}{2}\,(\vec r_i\times\hat{\sigma}_i)_z\, .
\label{ndop}
\end{equation}

The calculation of these non-diagonal
contributions to the magnetic moments calls for a specific
orbital wave
function model. 
Here, for simplicity, harmonic oscillator constituent
quark model wave functions are employed:
\begin{eqnarray}
\phi_{00}(\vec p;\omega)&=&{1\over (\omega^2\pi)^{3/4}}
\exp\{-\frac{p^{2}}{2\omega^{2}}\}\,\\
\phi_{1m}(\vec p;\omega)&=&\sqrt{2}\,{p_{m}\over \omega}\,
\phi_{00}(\vec p;\omega)\, .
\end{eqnarray}
Here $\phi_{00}(\vec{p};\omega)$ and $\phi_{1,m}(\vec{p};\omega)$ 
are the s-wave and p-wave
orbital wave functions of the constituent quarks,
respectively. The 
oscillator parameters of the $qqq$ and $qqqq\bar{q}$ components,
$\omega_{3}$ and $\omega_{5}$,
will in general be different.

The relation between the oscillator parameters $\omega_3$ and 
$\omega_5$ depends on the color dependence of the effective confining
interaction. If the confining interaction between two quarks
is pairwise with the color factor $\tilde \lambda_i·\cdot\tilde\lambda_j$,
where $\lambda^a_i$ is a color $SU(3)$ generator, the strength of the
pairwise confining interaction between the quarks in the
$qqqq$ subsystem is half of that between the quarks in a color
singlet $qqq$ triplet \cite{richard}. This would imply the relation:
\begin{equation}
\omega_5 = \sqrt{5/6}\,\omega_3.
\label{reln}
\end{equation}

The parameter $\omega_3$ may be determined by the nucleon 
radius as $\omega_3=1/\sqrt{<r^2>}$ or as half the splitting between
the nucleon and its lowest positive parity resonance. Both methods
yield the same value $\sim$ 246 MeV. The parameter $\omega_5$
may be set by the relation (\ref{reln}) or be treated as a
free phenomenological parameter. In ref.\cite{zou3} it was noted
that the best description of the extant empirical strangeness
form factors is obtained with $\omega_5\sim$ 1 GeV, which would imply
that the $qqqq\bar q$ component is very compact.  
 
\subsection{Magnetic moment expressions}

The magnetic moments of the octet
baryons are formed of diagonal matrix elements in the $qqq$
and $qqqq\bar q$ subspaces, respectively, and off-diagonal 
transition matrix elements of the form $qqq\rightarrow qqqq\bar q$
and $qqq\bar q\rightarrow qqq$. The former only depend on the
group theoretical factors, while the latter also depend on
the spatial wave function model.
The diagonal contributions to the octet magnetic moments
may be expressed in the form:
\begin{eqnarray}
\mu_{p}&=&P_{3q}\frac{M_{N}}{m}+P_{(p)s\bar{s}}(\frac{M_{N}}{6m}-
\frac{M_{N}}{6m_{s}})\, ,
\label{pmomD}\\
\mu_{n}&=&-P_{3q}\frac{2M_{N}}{3m}+P_{(n)u\bar{u}}
\frac{M_{N}}{3m}-P_{(n)s\bar{s}}
\frac{M_{N}}{6m_{s}}\, ,\label{nmomD}\\
\mu_{\Sigma^{+}}&=&P_{3q}(\frac{8M_{N}}{9m}+
\frac{M_{N}}{9m_{s}})+P_{(\Sigma^{+})s\bar{s}}
(\frac{2M_{N}}{9m}-\frac{2M_{N}}{9m_{s}})
+P_{(\Sigma^{+})d\bar{d}}(\frac{M_{N}}{18m}-
\frac{M_{N}}
{18m_{s}})\, ,\\
\mu_{\Sigma^{0}}&=&P_{3q}(\frac{2M_{N}}{9m}+\frac{M_{N}}{9m_{s}}
)+P_{(\Sigma^{0})s\bar{s}}
(\frac{M_{N}}{18m}-2\frac{M_{N}}{9m_{s}})-
P_{(\Sigma^{0})d\bar{d}}(\frac{M_{N}}{9m}+
\frac{M_{N}}
{18m_{s}})\, ,\nonumber\\
&&+P_{(\Sigma^{0})u\bar{u}}
(7\frac{M_{N}}{18m}-\frac{M_{N}}{18m_{s}})\, ,\\
\mu_{\Sigma^{-}}&=&-P_{3q}(\frac{4M_{N}}{9m}-\frac{M_{N}}{9m_{s}})
-P_{(\Sigma^{-})s\bar{s}}
(\frac{M_{N}}{9m}+\frac{2M_{N}}{9m_{s}})-
P_{(\Sigma^{-})u\bar{u}}(\frac{2M_{N}}{9m}+
\frac{M_{N}}{18m_{s}})\, ,\\
\mu_{\Sigma^{0}\rightarrow\Lambda}&=& -P_{3q}(\frac{M_{N}}{\sqrt{3}m})
+\frac{1}{4\sqrt{3}}P_{5q}\frac{M_{N}}{m}\, , \label{tmomD}\\
\mu_{\Xi^{0}}&=&P_{3q}(-\frac{2M_{N}}{9m}-\frac{4M_{N}}{9m_{s}})+
P_{(\Xi^{0})u\bar{u}}
(\frac{4M_{N}}{9m}-\frac{M_{N}}{9m_{s}})-
P_{(\Xi^{0})d\bar{d}}(\frac{M_{N}}{18m}+
\frac{M_{N}}{9m_{s}})\, ,\\
\mu_{\Xi^{-}}&=&P_{3q}(\frac{M_{N}}{9m}-\frac{4M_{N}}{9m_{s}})+
P_{(\Xi^{-})u\bar{u}}
(\frac{5M_{N}}{18m}-\frac{M_{N}}{9m_{s}})
-P_{(\Xi^{-})d\bar{d}}(\frac{2M_{N}}{9m}+
\frac{M_{N}}{9m_{s}})\, , \\
\mu_{\Lambda}&=&-P_{3q}\frac{M_{N}}{3m_{s}}
+P_{(\Lambda)u\bar{u}}(7\frac{M_{N}}{18m}
-\frac{M_{N}}{18m_{s}})-P_{(\Lambda)d\bar{d}}(\frac{M_{N}}{9m}+
\frac{M_{N}}{18m_{s}})\, .
\label{lambdam}
\end{eqnarray}
The factors
$P_{(B)q_i\bar{q_i}}$ are the probabilities of the
$qqqq_i\bar{q}_i$ components
in the baryon $B$. These are related to the 
corresponding amplitudes $A_i$ and the probability of the
$qqqq\bar{q}$ components (\ref{bwave}) as:
\begin{equation}
P_{(B)q_{i}\bar{q}_{i}}=P_{5q}A_{i}^{2}\, .
\end{equation}

The contributions to the octet magnetic moments from the off-diagonal 
matrix elements take the following forms in the harmonic
oscillator model:
\begin{eqnarray}
\mu_{p}&=&
-\frac{2\sqrt{3}}{9}F_{35}(P_{(p)s\bar{s}})
-\frac{2\sqrt{6}}{9}F_{35}(P_{(p)d\bar{d}})\ ,
\label{pmomND}\\
\mu_{n}&=&
-\frac{4\sqrt{6}}{9}F_{35}(P_{(n)u\bar{u}})
-\frac{2\sqrt{3}}{9}F_{35}(P_{(n)s\bar{s}})\, ,\\
\mu_{\Sigma^{+}}&=&
-\frac{2\sqrt{3}}{9}F_{35}(P_{(\Sigma^{+})d\bar{d}})
-\frac{2\sqrt{6}}{9}F_{35}(P_{(\Sigma^{+})s\bar{s}})\, ,\\
\mu_{\Sigma^{0}}&=&
+\frac{2\sqrt{6}}{18}F_{35}
(P_{(\Sigma^{0})s\bar{s}})-\frac{2\sqrt{6}}{9}F_{35}
(P_{(\Sigma^{0})u\bar{u}})+
\frac{\sqrt{6}}{9}F_{35}(P_{(\Sigma^{0})d\bar{d}})\, ,\\
\mu_{\Sigma^{-}}&=&
-\frac{2\sqrt{6}}{9}F_{35}(P_{(\Sigma^{-})s\bar{s}})
+\frac{4\sqrt{3}}{9}F_{35}
(P_{(\Sigma^{-})u\bar{u}})\, ,\\
\mu_{\Sigma^{0}\rightarrow\Lambda}&=&
-\frac{2\sqrt{3}}{3}F_{35}(P_{5q})
\label{tmom}\\
\mu_{\Xi^{0}}&=&
-\frac{2\sqrt{3}}{9}F_{35}(P_{(\Xi^{0})d\bar{d}},0)+\frac{4\sqrt{6}}{9}F_{35}
(P_{(\Xi^{0})u\bar{u}})\, ,\\
\mu_{\Xi^{-}}&=&
\frac{4\sqrt{3}}{9}F_{35}(P_{(\Xi^{-})u\bar{u}})-\frac{2\sqrt{6}}{9}F_{35}
(P_{(\Xi^{-})d\bar{d}})\, ,\\
\mu_{\Lambda}&=&
-\frac{1}{3}F_{35}(P_{5q})\, .
\end{eqnarray}
Here the phase factors for the off-diagonal
matrix element between the
$qqq$ and $qqqq\bar{q}$ components have been taken to be +1. 
The functions $F_{35}(P_{(B)q\bar{q}})$ above 
are defined as
\begin{equation}
F_{35}(P_{(B)q\bar{q}})=C_{35}\frac{M_{N}}{\omega_{5}}
\sqrt{P_{3q}P_{(B)q\bar{q}}}\, ,
\end{equation}
where the factor $C_{35}$,
\begin{equation}
C_{35}=(\frac{2\omega_{3}\omega_{5}}{\omega_{3}^{2}+
\omega_{5}^{2}})^{9/2}\, ,
\label{c35}
\end{equation}
is the overlap integral of the s-wave wave functions of the quarks
in the $qqq$ and $qqqq\bar{q}$ configurations.

Note that the $[22]$ symmetry of the flavor wave function rules
out any contribution to the magnetic moments of the $\Xi$ hyperons
from $s\bar s$ components. For the same reason there is no
contribution to the proton magnetic moment from $u\bar u$
nor to the neutron magnetic moment from $d\bar d$ components.

\subsection{Numerical results}

Here the factor $C_{35}$ (\ref{c35}) is treated as a free 
phenomenological parameter. It is found that addition of $qqqq\bar q$ 
admixtures reduces the deviation of the calculated values
from the experimental magnetic moment values only if the $qqqq\bar q$
component is much more compact than the $qqq$ component.
This will be shown by comparing the calculated values for
$C_{35}\sim 0.24$ and $C_{35} = 1$. The latter value implies equal
oscillator parameters in the $qqq$ and $qqqq\bar q$ systems,
while the former value corresponds to $\omega_5\sim 2.3\,\omega_3$.
The former value would imply that the radius of the $qqqq\bar q$
component is only $\sim$ 0.3 fm.

The other model parameters are the probabilities
of the  $qqqq\bar{q}$ components $P_{5q}$ and the constituent quark masses.
To reproduce the experimentally
measured values $\bar{d}-\bar{u}=0.12$ in the proton 
\cite {asymmetry} the $qqqq\bar q$ probability is set to
$P_{5q}=1-P_{3q}=0.18$. 
The constituent masses of the up and down quarks are set to be 
$m_{u}=m_{d}=m=274$ MeV and that of the
strange quark to be $m_{s}=419$ MeV. The oscillator parameter 
$\omega_{5}$ is taken to
have the values $0.57$ GeV, which corresponds to a compact 
$qqqq\bar q$ extension in the
baryon octets.  The corresponding
numerical results of the baryon octet magnetic moments are shown in
Table \ref{table4}.

\begin{table}[h]\caption{\footnotesize Magnetic moments of 
the baryon octet.
The column $qqq$ contains the results of the conventional 
quark model from Refs.\cite{riska} and 
the column Exp the experimental
data from \cite{pdg}.  The present results are listed in column $P_1$.
Columns D($qqq$) and D($P_1$) contain
the deviations of the calculated results from the data, respectively.}
\vspace{0.5cm}
\footnotesize
\begin{tabular}{cccccc}
\hline
Baryon          & Exp                 &        $qqq$             & 
   $P_1$       
  & D($qqq$)    &   D($P_1$)    \\\hline
p               & 2.79                &       2.76               &   2.72      
     &  1.1\%  &    2.5\%  \\
n               &-1.91                &      -1.84               &  -1.93   
        &  3.6\%  &    0.8\%  \\
$\Lambda$       &-0.61                &      -0.67               &  -0.61     
      &  9.8\%  &    0.0\%  \\
$\Sigma^{+}$    & 2.46                &       2.68               &   2.63      
     &  8.9\%  &    6.9\%  \\
$\Sigma^{0}$    & ?                   &       0.84               &   0.87     
    &   ?     &     ?     \\
$\Sigma^{-}$    &-1.16                &      -1.00               &  -1.11      
      & 13.7\%  &    4.3\%  \\
$\Xi^{0}$       &-1.25                &      -1.51               &  -1.21      
     & 20.8\%  &    3.0\%  \\
$\Xi^{-}$       &-0.65                &      -0.59               &  -0.58     
      &  9.2\%  &   10.3\%  \\
$\Sigma^0\rightarrow \Lambda$ & $|1.61|$ &     -1.59             &  -1.67      
      & 1.2\%   &    3.7\%  \\
\hline
\end{tabular}
\label{table4}
\end{table}

The results show that the average deviation from the empirical
magnetic moment values drops from $\sim$ 9\% to $\sim$ 4\%,
when the $qqqq\bar q$ admixture is included in the quark
model. The improvement is most notable in the case of the
$\Sigma^-$ and the $\Xi^o$ hyperons, which deviate most
notably from the empirical values in the conventional
$qqq$ model \cite{Franklin2}. For these the $qqq$ model 
results differ by 14\%
and 21\%, respectively,  from the experimental values. Inclusion
of the contributions of the $qqqq\bar{q}$ components reduce these
differences to $\sim$ 4\% and 3\%, respectively.

\begin{table}[h]\caption{\footnotesize Magnetic moments of the baryon 
octet.
The column Exp contains the experimental
data from \cite{pdg}.  The present results are listed in columns $P_{1}$, 
with model 
parameters $C_{35}=0.24, P_{3q}=0.82$; $P_{2}$, with $C_{35}=0.24,
 P_{3q}=0.90$; $P_{3}$, 
with $C_{35}=1, P_{3q}=0.82$; $P_{4}$, $C_{35}=1, P_{3q}=0.90$, 
 respectively.}
\vspace{0.5cm}
\footnotesize
\begin{tabular}{cccccc}
\hline
Baryon          & Exp                 &           
 $P_{1}$       
  & $P_{2}$  &$P_{3}$    &  $P_{4}$    \\\hline
p               & 2.79                                           &   2.72      
  &   3.01   &  2.40  &    2.76  \\
n               &-1.91                               &  -1.93  
     
 &   -2.12   &  -2.46  &    -2.54  \\
$\Lambda$       &-0.61                               &  -0.61      
  &  -0.68    &  -0.71  &    -0.80  \\
$\Sigma^{+}$    & 2.46                            &   2.63      
  &   2.90  &  2.31  &    2.65  \\
$\Sigma^{0}$    & ?                                 &   0.87      
  &    0.94  &   1.03     &     1.07    \\
$\Sigma^{-}$    &-1.16                               &  -1.11      
  &  -1.18    & -1.11  &    -1.18  \\
$\Xi^{0}$       &-1.25                               &  -1.21      
  &  -1.43   & -0.89  &    -1.18  \\
$\Xi^{-}$       &-0.65                              &  -0.58     
   &  -0.60   &  -0.58  &   -0.60  \\
$\Sigma^0\rightarrow \Lambda$ & $|1.61|$             &  -1.67      
  &  -1.84    & -2.12   &   -2.19  \\
\hline
\end{tabular}
\label{table5}
\end{table}

In Table \ref{table5} the calculated magnetic moments are shown for both 
$P_{5q}=0.1$ and $P_{5q}= 0.18$ and for both $C_{35}=0.24$ and  $C_{35}=1.0$.
The results indicate a clear preference for the smaller value
of $C_{35}$ and the larger value of $P_{5q}$.

\section{Decuplet magnetic moments}
\label{sec:4}

In the case of the decuplet baryons the lowest energy
$qqqq\bar q$ configuration is expected to be that, for which the
$qqqq$ subsystem is 
assumed to have the $[4]_{FS}[31]_F[31]_S$
mixed flavor-spin symmetry.
The corresponding flavor wave functions are listed in
table \ref{table2}, and the combined flavor-spin wave
function in eq. (\ref{lowestb}). 
This wave function, and the corresponding conventional
wave function in the $qqq$ quark model lead to the following 
diagonal matrix element contribution to the magnetic moment
values for the $\Delta^{++}$ and the $\Omega^-$ members of
the baryon decuplet for which experimental data exist:  
\begin{eqnarray}
\mu_{\Delta^{++}}&=&P_{3q}\frac{2M_{N}}{m}+P_{(\Delta^{++})d\bar{d}}
\frac{35M_{N}}{24m}+P_{(\Delta^{++})s\bar{s}}
(\frac{13M_{N}}{12m}+\frac{3M_{N}}{8m_{s}})\, ,\\
\mu_{\Omega^{-}}&=&-P_{3q}\frac{M_{N}}{m_{s}}+P_{(\Omega^{-})u\bar{u}}
(-\frac{13M_{N}}{24m_{s}}-\frac{3M_{N}}{4m_{s}})
+P_{(\Omega^{-})d\bar{d}}(\frac{3M_{N}}{8m}-\frac{13M_{N}}{24m_{s}})
\, .
\label{DeltaD}
\end{eqnarray}
The off-diagonal $qqq\rightarrow qqqq\bar q$ matrix
element contributions are
the following:
\begin{eqnarray}
\mu_{\Delta^{++}}&=&-\frac{\sqrt{3}}{12}F_{35}(P_{(\Delta^{++})d\bar{d}},0)
-\frac{\sqrt{3}}{12}F_{35}(P_{(\Delta^{++})s\bar{s}},0)\, ,\\
\mu_{\Omega^{-}}&=&\frac{\sqrt{3}}{6}F_{35}(P_{(\Omega^{-})u\bar{u}},0)
-\frac{\sqrt{3}}{12}F_{35}(P_{(\Omega^{-})d\bar{d}},0)\, .
\label{omegaND}
\end{eqnarray}
Note that the $qqqq$ subsystem of the 
$qqqq\bar q$ component in baryon decuplet 
can have both total angular momentum $J=1$ and $J=2$. 
Here only the contribution from $J=1$, which, with 
40\% and 20\% proportion of the $qqqq\bar q$ component in baryon deculplet, 
leads to the 
values in table \ref{table6}.

\begin{table}[h]\caption{\footnotesize The magnetic moments of the baryon
decuplet.
The column $qqq$
contains $qqq$ model results \cite{riska}
 and 
the column Exp the experimental
data from \cite{pdg}.  The present results are listed in columns $P_{1}$
 with model parameters
$C_{35}=0.24, P_{3q}=0.60$; $P_{2}$ with $C_{35}=1, P_{3q}=0.60$; 
$P_{3}$ with $C_{35}=0.24, P_{3q}=0.80$; $P_{4}$ with
$C_{35}=1, P_{3q}=0.80$.
}
\vspace{0.5cm}
\footnotesize
\begin{tabular}{ccccccc}
\hline
Baryon          & Exp                 &        CQ                &   
  $P_{1}$       
 & $P_{2}$        & $P_{3}$             & $P_{4}$  \\\hline
$\Delta^{++}$       &3.7-7.5                &      5.52              & 4.98  
& 4.94             &   5.87       & 5.84\\
$\Omega^{-}$       &-2.02                &      -2.01               & -2.02 
& -2.00          & -2.11        &-2.09   \\
\hline
\end{tabular}
\label{table6}
\end{table}

Note that the $[31]$ symmetry of the flavor wave function rules
out any contribution to the magnetic moments of the $\Omega^-$ hyperon
from $s\bar s$ components. For the same reason there is no
contribution to the $\Delta^{++}$ magnetic moment from $u\bar u$
components.

The values in table \ref{table6} shows that the inclusion of  
the $qqqq\bar q$ components leads to 
improved agreement with the empirical value of the
the empirically well determined magnetic moment 
of $\Omega^{-}$  
only if the $qqqq\bar q$ system is again very compact.
In the case of the magnetic moment of the $\Delta^{++}$
the uncertainty range of the empirical values is too
wide to allow any definite conclusion. 

\section{Strangeness magnetic moments}
\label{sec:5}

The contribution to the strangeness magnetic moment of the
proton can be inferred directly from eqs. (\ref{pmomD})
and (\ref{pmomND}). With $P_{5q}=0.18$ $SU(3)$ symmetry
suggests $P_{s\bar{s}}\sim 0.06$. With this value the
strangeness magnetic moment of the proton is $\mu_{s}=0.17$,
which falls well within the ranges
of the empirical data ($0.37\pm 0.2\pm 0.26\pm 0.07$) given by the SAMPLE
experiment\cite{sample1} and of the combined value from all the presently
completed experiments ($0.28\pm 0.20$) \cite{Milos}. 

The contributions from $s\bar s$ seaquark configurations to any
of the octet and decuplet magnetic moments can be directly
calculated from the terms in eqs.(\ref{pmomD})-(\ref{lambdam}) 
and (34), (36) that are proportional
to $P_{(B)(s\bar s)}$ and inversely proportional to the
constituent mass $m_s$ of the strange quark. From these
it emerges that the strangeness magnetic moments
of the proton $\mu_s(P)$ and the neutron $\mu_s(N)$ are
equal as required by $SU(3)$ flavor symmetry. For the
same reason $\mu_s(\Sigma^+)=\mu_s(\Sigma^-)$.

As noted above there are no $s\bar s$
contributions to the magnetic moments of the
$\Xi$ hyperons because of the restriction to $[22]_F$,
nor to that of the $\Omega^-$ hyperon because of 
the restriction to $[31]_F$ symmetry. Such contributions
are nevertheless possible in the case of the
$\Xi$ hyperons if the flavor symmetry is $[31]_F$ and
in the case of the $\Omega^-$ hyperon if the flavor
symmetry is $[4]_F$ instead. Those configurations are
however expected to be energetically unfavorable.

\section{Light flavor seaquark magnetic moments}
\label{sec:6}

The contributions to the magnetic moments from
one $u\bar u$ seaquark component can be derived from
those terms in the expressions (\ref{pmomD})-(\ref{lambdam}) 
and (\ref{DeltaD}),(\ref{omegaND}), which are proportional
to $P_{(B)(u\bar u)}$ and inversely proportional to the
light flavor constituent mass $m$. In the case of the
octet baryons the resulting expressions naturally
only apply to the symmetry $qqqq$ configuration
$[4]_{FS}[22]_F[22]_S$. In the case of the decuplet
baryons the corresponding expressions (\ref{DeltaD}),(\ref{omegaND})
are restricted to the symmetry configuration
$[4]_{FS}[31]_F[31]_S$.

These two symmetries imply that in the case of the proton
only $d\bar d$ contributions, but no $u\bar u$ contributions
appear. This is in line with empirical observation
that the $\bar d$ contributions are larger than the
$\bar u$ contribution \cite{garvey}. Similarly there are
no $d\bar d$ contributions to the magnetic moment of the
neutron. 

\section{Conclusions}
\label{sec:7}

Here the magnetic moment contributions from $qqqq\bar q$
admixtures in the baryon octet and decuplet have been
derived within the framework of the flavor SU(3) symmetry.
In addition explicit expressions for all the possible
flavor wave functions for $qqqq\bar q$ systems with
completely symmetric combinations of flavor and spin
wave functions are given. The calculated magnetic moments
include the contributions from the $qqq$ components and the 
$qqqq\bar q$ components with both light and strange $q\bar q$ pairs. 
The magnetic moment expressions readily allow separation of
the strangeness components from $s\bar s$ pairs as well
as individual components from $u\bar u$ and $d\bar d$
pairs.

If the $qqqq\bar q$ components are more compact than the
$qqq$ components of the wave functions an improved 
description of the 
experimental magnetic moments is possible, with appropriately
chosen model parameter values. With a $qqqq\bar q$ probability
in the range $\sim$ 10 - 20\% the $qqqq\bar q$ 
components lead to small corrections to the magnetic moment 
values given by the conventional $qqq$ model apart from the
$\Sigma^{-}$ and $\Xi^{0}$ hyperons, where these corrections
are large and notably improve the agreement with the
empirical values.

The restriction in the calculation of the baryon
octet magnetic moments to the
configuration with
flavor-spin symmetry $[4]_{FS}[22]_F[22]_S$ given in Table
\ref{table3}, is motivated on the one hand by its expected 
low energy and on the other hand by the indications of the
experimentally observed positive sign of the strangeness
magnetic moment of the proton, which is best described
by this configuration \cite{zou1,zou2,zou3}. 
The configuration with
$[4]_{FS}[31]_F[31]_S$ symmetry given in Table \ref{table2} is
expected to have the next lowest energy and to give an
at most very insignificant 
contribution to the baryon octet magnetic moments.
In the case of the decuplet this configuration is, however, 
expected to have the lowest energy as the configuration
$[4]_{FS}[22]_F[22]_S$ cannot contribute. 
The contribution of admixtures of $qqqq\bar q$ components
with the flavor-spin symmetry $[4]_{FS}[31]_F[31]_S$ were
found to be very small in the case of the $\Omega^-$.

It is of course not obvious that those 
$qqqq\bar q$ configurations that have the lowest energy
for a given hyperfine interaction model should have the
largest probability in the nucleons. The main terms
should be expected to be those with the strongest coupling
to the $qqq$ component. This coupling depends both on the
confining interaction in the transition amplitude and 
(inversely) on the difference in energy from the rest energy 
of the nucleon. 

It should be noted that the $qqqq\bar q$ components here,
for which the antiquark is in its ground state, do not 
correspond to the pion contributions considered e.g. in
refs.\cite{Franklin1,Franklin2}. In the quark model
those pion contributions correspond to $qqqq\bar q$ 
configurations, in which the antiquark is in the $P-$state,
and the $qqqq$ system is in the ground state. The
present approach is motivated by the empirical 
observation that the strangeness magnetic moment of the
proton is positive, which only can be described
by $qqqq\bar q$ configurations with the antiquark
in the ground state \cite{zou1,zou2,zou3}.

\begin{acknowledgments}

C. S. An acknowledges the hospitality of the Helsinki Institute of Physics
 during the course of this
work. Research supported in part by the National
Natural Science Foundation of China under grants Nos.10225525 \& 10435080.

\end{acknowledgments}

\end{document}